\documentclass[draft,jgrga]{agutex}

 \usepackage{lineno}
 \linenumbers*[1]

\usepackage{graphics}
\usepackage{epsfig}
\usepackage{url}
 \setkeys{Gin}{draft=false}
\authorrunninghead{Kronberg et al.}

\titlerunninghead{Oxygen abundance in the magnetosphere}

\authoraddr{E. A. Kronberg,
Max-Planck Institute for Solar System Research, Katlenburg-Lindau, Germany
(kronberg@mps.mpg.de)}

\authoraddr{S. E. Haaland,
Max-Planck Institute for Solar System Research, Katlenburg-Lindau, Germany.\\
Also at Department of Physics and Technology, University of Bergen, Norway.
(stein.haaland@ift.uib.no)}

\authoraddr{P. W. Daly,
Max-Planck Institute for Solar System Research, Katlenburg-Lindau, Germany
(daly@mps.mpg.de)}

\authoraddr{E. E. Grigorenko, Space Research Institute, Russian Academy of Sciences,
Moscow, Russia (elenagrigorenko2003@yahoo.com)}

\authoraddr{L. M. Kistler,
University of New Hampshire, Durham, NH, USA
(Lynn.Kistler@unh.edu)}

\authoraddr{M. Fr\"anz,
Max-Planck Institute for Solar System Research, Katlenburg-Lindau, Germany
(fraenz@mps.mpg.de)}

\authoraddr{I. Dandouras,
IRAP, University of Toulouse, UPS-OMP, CNRS, Toulouse, France
(Iannis.Dandouras@irap.omp.eu)}

\begin{document}

\title{Oxygen and hydrogen ion abundance in the near-Earth magnetosphere: Statistical results
on the response to the geomagnetic and solar wind activity conditions}

\authors{
E. A. Kronberg  \altaffilmark{1},
S. E. Haaland \altaffilmark{1,2},
P. W. Daly \altaffilmark{1},
E. E. Grigorenko \altaffilmark{3},
L. M. Kistler \altaffilmark{4},
M. Fr{\"a}nz  \altaffilmark{1},
 and
I. Dandouras \altaffilmark{5}}

\altaffiltext{1}{Max Planck Institute for Solar System Research,
Katlenburg-Lindau, Germany}

\altaffiltext{2}{Department of Physics and Technology,
University of Bergen, Norway}

\altaffiltext{3}{Space Research Institute, Russian Academy of Sciences, Moscow, Russia}

\altaffiltext{4}{University of New Hampshire, Durham, NH, USA}

\altaffiltext{5}{Institut de Recherche en Astrophysique et Plan\'{e}tologie,
University of Toulouse, UPS-OMP, CNRS, Toulouse, France}

%
%

\begin{abstract}
The composition of ions plays a crucial role for the fundamental plasma
 properties in the terrestrial magnetosphere. We investigate the
 oxygen-to-hydrogen ratio in the near-Earth magnetosphere from -10 \,R$_{\rm E}$\
 $<X_{GSE}<$ 10 \,R$_{\rm E}$. The results are based on seven years of ion flux
 measurements in the energy range $\sim$10 keV to $\sim$955 keV from the RAPID
 and CIS instruments on board the Cluster satellites. We find that (1)
 hydrogen ions at $\sim$10 keV show only a slight correlation with the geomagnetic conditions and
interplanetary magnetic field changes. They are best correlated with the
solar wind dynamic pressure and density, which is an expected effect of the
magnetospheric compression; (2) $\sim$10 keV O$^+$ ion intensities are more
strongly affected during disturbed phase of a geomagnetic storm or substorm than $>$274 keV O$^+$
ion intensities, relative to the corresponding hydrogen intensities; (3) In
contrast to $\sim$10 keV ions, the $>$274 keV O$^+$ ions show the strongest
acceleration during growth phase and not during the expansion phase itself.
This suggests a connection between the energy input to the magnetosphere and
the effective energization of energetic ions during growth phase; (4) The
ratio between quiet and disturbed times for the intensities of ion ionospheric outflow is
similar to those observed in the near-Earth magnetosphere at $>$274 keV.
Therefore, the increase of the energetic ion intensity during disturbed time
is more likely due to the intensification than to the more
effective acceleration of the ionospheric source. In conclusion, the energization process in the
near-Earth magnetosphere is mass dependent and it is more effective for the
heavier ions.

\end{abstract}

\begin{article}
\section{Introduction}
\label{sec:intro}

There are two major sources for the ion population in Earth's magnetosphere:
the solar wind and the ionosphere.

Hydrogen, helium and oxygen are the most abundant elements in the solar wind.
According to measurements by SWICS on Ulysses, the oxygen abundance (H/O
ratio) varies significantly between fast and slow solar winds, with mean
values of 1500 and 2300, respectively \citep[e.g.,][]{Steiger10}. A direct
entry of solar wind plasma into the magnetosphere is believed to occur mainly
when the interplanetary magnetic field (IMF) is oriented southward. This
leads to reconnection first at the dayside magnetopause and then in the
magnetotail. Also, northward IMF results in reconnection at southern and
northern high latitudes, facilitating the entry of solar wind plasma.

The main source of oxygen in the magnetosphere is believed to be the
terrestrial ionosphere
\citep{yau84a,Lockwood85,gloeck85a,Chappell87,yau97,Chappell00,cully03,hudd05a,
Moore07,Kita2010}, where oxygen ions are accelerated upward by electric
fields parallel to the background magnetic field and pressure gradients. The
ionospheric ions are either accelerated in the cusp/cleft region and then
convected across the polar cap into the lobe \citep{Lockwood85,Kistler2010b},
or they come from the nightside aurora, which provides a fast feeding of
O$^+$ ions in the near-Earth plasma sheet during the substorm expansion phase
\citep{yau85,Daglis96}. The kinetic energy of upflowing ions is typically a
few tens of eV \citep{yau97}, whereas typical plasma sheet energies are much
higher \citep[e.g.,][]{lenn86a}. Centrifugal forces accelerate particles as
they travel tailward along field lines \citep[e.g.,][]
{clad86a,nils08a,nils10a}, but the most significant energization will take
place after the particles enter the plasma sheet.

The accelerated plasma sheet particles are convected Earthward and populate
the ring current which influences the transport in the radiation belts. The
radiation belts are a key region for space weather studies, as their dynamics
can induce damage to satellites and other space weather effects. Forecasting
models based on radiation belt dynamics include the dynamics of the ring
current. Additional ionospheric particles affect the background magnetic
field, densities, temperature and convection electric field, and therefore
lead to, e.g., changes in the ring current \citep{Welling11}. It is
challenging to include the heavy ion contribution in numerical studies of the
ring current \citep[e.g.,][]{Glocer09,Welling11} or radiation belt models.
Knowledge about the ion composition and its dependence on solar wind and
magnetospheric conditions is important for defining the boundary conditions
in such simulations.

A number of earlier studies have demonstrated that the abundance of oxygen in
the ring current region can change significantly with geomagnetic activity.
\citet{Sharp81} have reported measurements of oxygen in the energy range
0--17 keV/e at distances between 11 and 23 \,R$_{\rm E}$\ downtail by the ISEE 1
spacecraft. \citet{gloeck85a} analyzed for the first time the ion composition
measurements of the ring current during the storm time using AMPTE/CCE
spacecraft data in the energy range $\sim$1 to 300 keV/e at L values between
$\sim$3 and 6.5. They suggested that large changes in the O$^+$ number and energy
densities implied an injection of energetic ionospheric ions.  \citet{hami88a},
using AMPTE particle measurements, found an event in which the O$^+$
dominates H$^+$ in the ring current during a large magnetic storm.  A later
study by \citet{Daglis00} established a connection between O$^+$ (50-426 keV,
measured by CRRES) enhancements during storms and substorm activities in the
inner magnetosphere at L$\sim$7. Another, more extensive study by \citet{Fu02},
using CRRES observations, has shown that not all injection events have the same
ion composition. Their statistical results demonstrated that about 73$\%$ of all the
events were rich in oxygen ions and the rest have very low admixtures of oxygen.
Events without oxygen ions were found to be mainly correlated with weak geomagnetic
activity conditions.

The abundance of oxygen in the magnetotail's plasma sheet can vary
significantly with geomagnetic activity. \citet{moeb87} showed that the
spectra harden in the magnetotail's plasma sheet during substorm, with a
larger energy increase for O$^+$ than for H$^+$, i.e., the O$^+$/H$^+$ ratio
also increases with energy. For this case study they used observations from
AMPTE/IRM which resolved the ion composition of the suprathermal population
(10-230 keV/e). \citet{Kistler90} used AMPTE/IRM observations to compare
spectral changes at 15--19 \,R$_{\rm E}$\ (tail) and 7--9 \,R$_{\rm E}$\ (inner magnetosphere,
night side). They found that the spectral changes were very similar at both
locations. \citet{Kistler94} investigated how the flux of the energetic
particles observed by the AMPTE/IRM decreases after the initial particle
``injection''. These observations show that the decay rate is organized by
total energy, energy per charge and velocity. In a statistical study,
\citet{nose2009a} used 16 years of Geotail data in order to look at the
suprathermal ion composition in the plasma sheet in the range
$-100R_E<X_{GSM}<-8R_E$. They investigated the correlation between the
O$^+$/H$^+$ ratio and solar activity using the F10.7 index and with the
geomagnetic activity using the Kp index. They found that physical processes
in the plasma sheet are expected to be much different during solar minimum
and solar maximum, because the Alfv\'{e}n velocity changes significantly
between these two solar activity regimes. More recently, ion composition data
from the CIS/CODIF instrument on the Cluster spacecraft were used to study
O$^+$ and H$^+$ bulk content (0--40 keV/e) in the plasma sheet within 15--19
\,R$_{\rm E}$\ as a function of solar activity and geomagnetic activity
\citep{Mouikis10}. They found that the O$^+$ density were strongly dependent on
these parameters.

In this paper, we present a comprehensive study of the O$^+$ and H$^+$
abundance in the plasma sheet of the near-Earth magnetosphere (from -10 \,R$_{\rm E}$\
$<X_{GSE}<$ 10 \,R$_{\rm E}$). The data are obtained from seven years of particle
measurements by the Cluster satellites. We compare the characteristic behavior of the ions at
$\sim$10 keV and at 274--955 keV to different solar wind and geospace
activity factors.

The novelty of this study is a combination of (a) the extensive spatial
region coverage of the near-Earth magnetosphere, (many previous studies either
focused on the magnetotail region or were mainly case studies in the
near-Earth region only); (b) energy range, as no extensive statistical study has
been done so far with energies up to $\sim$955 keV; (c) a comparative analysis
of ion intensities at different energies, namely at $\sim$10 keV and from 274
keV to $\sim$955 keV; (d) the advantage of having more accurate parameters of the
solar-terrestrial coupling which became available during the last 15 years.

We can for the first time investigate the details of how the near-Earth
hydrogen and oxygen populations depend on the geomagnetic and solar wind
activity.

The paper is organized as follows: In Section \ref{sec:data}, we give a brief
overview of the Cluster spacecraft, the RAPID and CIS ions spectrometers and
their data products. In Section \ref{sec:method}, we explain how we process
the data and derive the O$^+$/H$^+$ ratio. Section \ref{sec:results} presents
the observations and shows how the O$^+$ and H$^+$ abundance varies with
different parameters controlling the dynamical state of the magnetosphere
such as solar wind conditions and also geomagnetic disturbance level. Section
\ref{sec:discussion} discusses the results and related physical processes.
Section \ref{sec:summary} summarizes the observations.

\section{Data and instrumentation}
\label{sec:data}

The results presented in this study are primarily based on in-situ
measurements from the Cluster quartet of spacecraft. The Cluster mission
comprises four identical spacecraft flying in a tetrahedron-like formation.
More information about the Cluster mission and instrumentation is given in
\citet{esco97a}. In this paper we have used data from the SC4 spacecraft, since
this gives the best data return for our purpose.

\begin{figure}
    {\includegraphics[width=130mm]{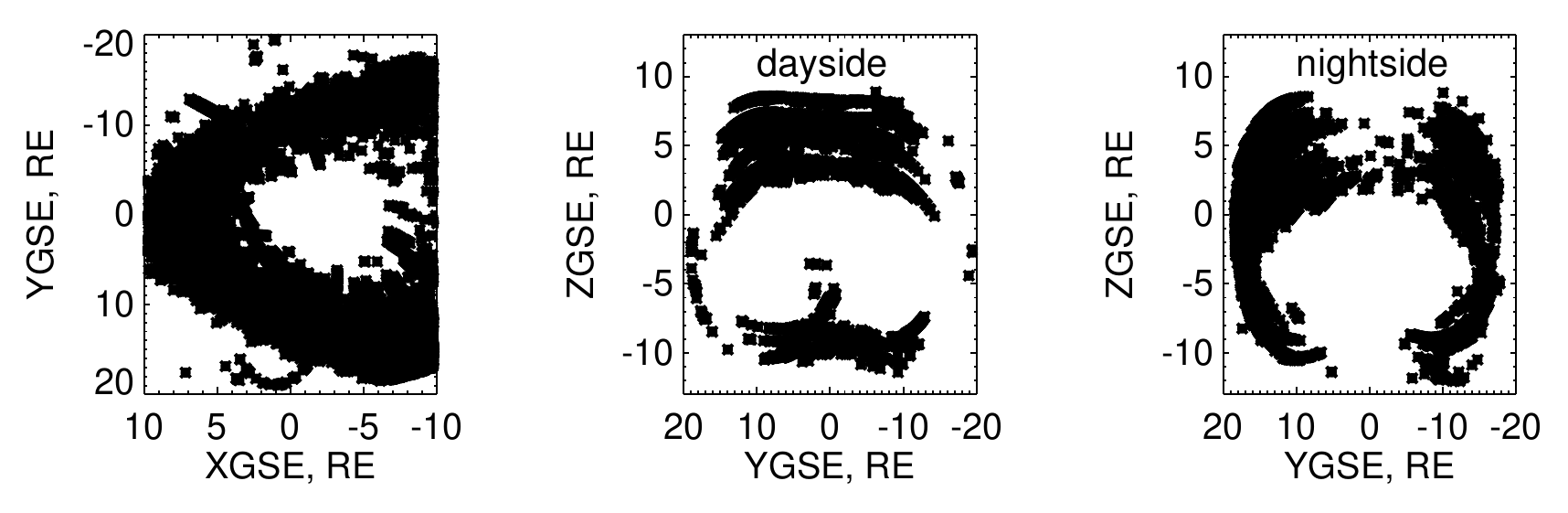}
  \caption{ Maps of the location of the energetic oxygen observations, from
  left in $XY_{GSE}$, $YZ_{GSE}$ dayside ($X_{GSE}>0$) and $YZ_{GSE}$ nightside ($X_{GSE}\le0$) planes.} \label{fig:mapYZ}}
\end{figure}

For the period covered by this study Cluster was in a nearly 90$^\circ$
inclination elliptical orbit with apogee around 19.6 \,R$_{\rm E}$\ and perigee around
4 \,R$_{\rm E}$. Due to spacecraft orbits in the near-Earth region our observations mainly cover
the dawn and dusk flanks and the dayside plasmasheet, mainly the north part, see
Figure \ref{fig:mapYZ}. This Figure shows maps of the energetic O$^+$
observations, projected into $XY_{GSE}$, $YZ_{GSE}$ dayside and $YZ_{GSE}$ nightside
planes, respectively. This coverage allows us to study the transport of ions from the tail
region around the Earth.

In addition to the in-situ Cluster observations, we also use solar wind
magnetic field and plasma data and geomagnetic disturbance indices to check
dependencies and correlations.

In this study, we use the AE and Dst indices and not the Kp index as e.g.
\citet{nose2009a} or \citet{Mouikis10} because they are more precise (the Kp
index has a 3 hours cadence, while the AE and Dst indices have higher time
resolution) and have a more direct physical meaning. The Dst index primarily
reflects slow (hours) magnetic perturbations caused mainly by ring current
enhancements, whereas the AE index reflects short time (minutes) perturbations at high
latitudes caused by variations in field-aligned current due to plasma sheet
processes such as bursty bulk flow activity.

\subsection{Cluster particle measurements}
\label{sec:instrumentation}

We use ion measurements from two different sensors onboard Cluster.
Energies above 30 keV are covered by RAPID: the Research with Adaptive Particle
Imaging Detector \citep[see][]{Wilken01}, whereas the lower part of
the energy range (energies below 40 keV) is covered by CIS: the Cluster Ion
Spectroscopy instrument \citep[see][]{reme01}.

\subsubsection{Energetic ion fluxes from RAPID }
\label{sec:rapid}

The RAPID Imaging Ion Mass Spectrometer (IIMS) consists of three identical
detector heads and use a combination of time-of-flight and energy
measurements to classify and bin the incident particles. The three sensors
cover 180$^\circ$ in polar angle, and by utilizing the spacecraft spin,
RAPID/IIMS provides the distribution of ions with complete coverage
of the unit sphere in phase space. Due to a degradation of the
central heads on the ion sensor on all spacecraft during the early phase of
the mission, only a limited part of the full 3D distribution is available
during later stages, however. Since our purpose is to investigate the statistical
abundance of two distinct ion species in a particular plasma region, we are
not dependent on any 3D abilities, but integrate over all directions to get
an omnidirectional flux. Except for extremely anisotropic distributions, the
lack of sensitivity in the central sensor does not play any major role
\citep{Kron10}.

RAPID/IIMS allows for discrimination of various ion species by utilizing
time-of-flight and energy information of each incident
particle. The resolution of the instrument makes it possible to group
ions into three categories; protons (H$^+$ atomic mass unit 1), helium (He, atomic
mass units 2-4) and a common group of heavy ions (atomic mass units 14-16).
This latter group is dominated by oxygen, but the mass resolution does not allow for
distinction between oxygen, carbon and nitrogen. This group is therefore
commonly refereed to as the CNO group. The energy ranges are 28 keV -- 4 MeV
for protons, 137 keV -- 3.8 MeV for helium and 274 keV -- 4 MeV for the CNO group.
No information about charge state is possible. For simplicity, we will simply
refer to the ratio $\mbox{O}^+/\mbox{H}^+$ in case of RAPID data, although, we mean the
CNO mass group and hydrogen, respectively.

The time resolution of the RAPID omnidirectional data is 1 spin ($\sim$4 seconds),
but since the count rates are often very low, we have used 1 minute
averages in this study.

\subsubsection{Moments and particle measurements from CIS}
\label{sec:cis}

CIS consists of two sensors. In this paper we use data from the COmposition
and DIstribution Function (CODIF) instrument to calculate an omnidirectional
flux as done for RAPID. The energy range for CIS is 0.03-40 keV per charge. In
addition to the omnidirectional proton and O$^+$ fluxes, we also utilize
plasma moments from the CODIF sensor.

As for RAPID we use 1 minute averages and put them on the same timeline as
the rest of the data set. Plasma moments are used to identify the plasma
region and to filter the data. The CIS instrument has several sensitivity
modes, automatically switched in-flight in order to achieve optimal
sensitivity and resolution for a particular plasma region. To avoid errors
introduced by mode shifting and non-optimal modes we only use data from the
so-called magnetospheric mode 13 \citep[see details in][]{reme01,dand05a}.

O$^+$ measurements by CIS can be slightly contaminated by protons (the major
species). The contamination of H$^+$ into O$^+$ is about 0.1-0.3$\%$
(depending on energy and on a particular mode) for low count rates (e.g. in
the magnetosphere) for $\sim$10 keV ions. In order to compensate for this
contamination of oxygen by protons during active time periods, we calculate
the oxygen intensities as follows: $j^{O^+}=j^{O^+}_m-j^{H^+}\cdot0.001,$
where $j^{O^+}_m$ and $j^{H^+}$ are the measured oxygen and proton
intensities, respectively.

\section{Methodology}
\label{sec:method}

\subsection{Calculation of the $\mbox{O}^+/\mbox{H}^+$ ratio}
\label{sec:ratio}

Our purpose is to compare $\mbox{O}^+/\mbox{H}^+$, namely, the ratio between intensities
which are proportional to the number fluxes of O$^+$ and H$^+$ [1/cm$^2$ sr s
keV] in two energy ranges. This ratio is equal to the ratio between integral
fluxes if the energy range is identical for the two species. To be able to
identify any spectral changes, we use two different energy ranges, $\sim$10
keV (8.1 to 10.3 keV) and the energy band 274--955 keV. These
energy bands are somewhat arbitrarily chosen, and are a compromise between
the energy binning of the two instruments and sufficient statistics. We take
a wide energy band for the energetic ions since the measurements of ions at
these energies have relatively low count rates and wide energy channel
divisions. As we do not consider details of any particular acceleration
process but only the response of the particle population to the different
geomagnetic and solar wind activity, the use of the wide energy range of the
energetic ions is justified in this study.

For the CIS instrument, the ratio between the number fluxes of O$^+$ and H$^+$
is based on the same energy range; since the energy channels are similar for
protons and oxygen. For RAPID, however, the energy binning for O$^+$ and
H$^+$ have different thresholds, see energy ranges listed in Table
\ref{tab:energies}. The details on how the RAPID energy channels were
rebinned for the most effective intensities comparison are described in the
Appendix. These RAPID intensities ($J^{H^+}$ and $J^{O^+}$) are used in the
current study.

\begin{table}
\caption{Lower thresholds of energy channels for protons and oxygen for RAPID. Note that energy channels for H$^+$ and
O$^+$ are not overlapping. To make the O$^+$/H$^+$ ratio meaningful, we therefore fit a
spectrum and derive an additional H$^+$ energy channel for RAPID, see Appendix.}
\label{tab:energies}
\begin{tabular}{cll}
Channel &      Energy range H$^+$, keV&    Energy range O$^+$, keV\\
\hline
1&        28-64\textsuperscript{a}&       84\textsuperscript{b}\\
2&        75&      274\\
3&        92&      414\\
4&       160&      498\\
5&       374&      638\\
6&       962&      948\\
7&      1885&     1414\\
8&      ---\textsuperscript{c}&     2539\\
Upper& 4007&      4046\\
\hline  \\
\end{tabular}
{\textsuperscript{a} Energy gap between hydrogen 1 $\&$ 2}\\
{\textsuperscript{b} CNO channel 1 contaminated at
       times, suppressed}\\
       {\textsuperscript{c} 8th channel not accessible for hydrogen}\\
\end{table}

\subsection{Construction of plots}
\label{sec:ratio}

Figures \ref{fig:dst} to \ref{fig:swpressure} show median intensities for protons and oxygens in two
different energy bands versus the respective geomagnetic or solar wind
parameter. The plots are constructed from data accumulated in the plasma
sheet region, defined by plasma beta value in the range 0.1--10
\citep{baum89a}. We have excluded observations where the
magnetic field component $B_z$ was negative in order to avoid the cusp region. Furthermore, we
limit our study to the near-Earth magnetosphere, i.e. distances
$-10$\,R$_{\rm E}$$<X_{GSE}<10$\,R$_{\rm E}$. We also exclude data from radial distances
$<6$\,R$_{\rm E}$, as the sensors can be affected by pileup effects from both penetrating
particles (e.g. during solar proton events) and high count rates.

To minimize the effect of skewed distribution, we use median rather than mean
values of the distribution functions. For the same reason, statistical
spreads are given as median absolute deviations (MAD), \citep[see
e.g.,][]{Venables99}. Median value and MAD are less sensitive to non-normal
distribution functions than the mean value and standard deviation. In order
to determine whether differences between values are significant, confidence
interval (CI) error bars are given in Figures \ref{fig:dst} to
\ref{fig:swpressure}: $CI=SE\cdot t_{n-1}$, where $t_{n-1}$ is the
$t$-distribution with $n$ degrees of freedom, $SE=\sigma/\sqrt{n}$ is
standard error, where $\sigma,$ the standard deviation, was calculated in
this case as: $\sigma=1.4826\cdot MAD$ \citep{Huber81}.

The O$^+$/H$^+$ ratio is defined as the ratio between median values of the
O$^+$ and H$^+$ intensities. Correspondingly, errors for the O$^+$/H$^+$ ratio are
calculated as: $CI_{O^+/H^+} = \sqrt{(CI_{O^+}/O^+)^2+(CI_{H^+}/H^+)^2}\cdot
\mbox{O}^+/\mbox{H}^+$. Each value shown in the Figures contains at least 100 measurements.

The histogram ranges for Dst, AE, solar wind density and solar wind pressure
are defined as follows. The distributions of these parameters are skewed. In
order to avoid outliers, which would take attention away from the more
typical values we (1) compute quartiles of the distribution; (2) define
interquartile range (IQR) between first quartile, Q1, and third quartile,
IQR=Q3-Q1; (3) use the 1.5$\cdot$IQR rule for outliers, which approximately
corresponds to 3$\sigma$. This will lead to the histogram ranges between
Q1-1.5$\cdot$IQR and Q3+1.5$\cdot$IQR. The behavior of the oxygen and proton
intensities are rather complicated at values which are larger than $\pm
3\sigma.$ As an exception we extended the range for the Dst from -70 nT to
-100 nT, in order to cover the whole range for moderated storms.

\section{Results}
\label{sec:results}

Below, we present and discuss how the O$^+$, H$^+$ intensities and
O$^+$/H$^+$ ratio depend on respective parameters of the solar-terrestrial
coupling in more detail.

\subsection{Response to magnetic activity}

\subsubsection{Geomagnetic storms}\label{sec:Dst}

The interaction between a solar wind shock wave and/or an interplanetary magnetic
cloud with the terrestrial magnetic field can lead to the phenomena called
magnetic storms during which the horizontal component of the Earth's magnetic
field decreases dramatically. This decrease is reflected by the Dst index.

\begin{figure}
  {\includegraphics[width=150mm]{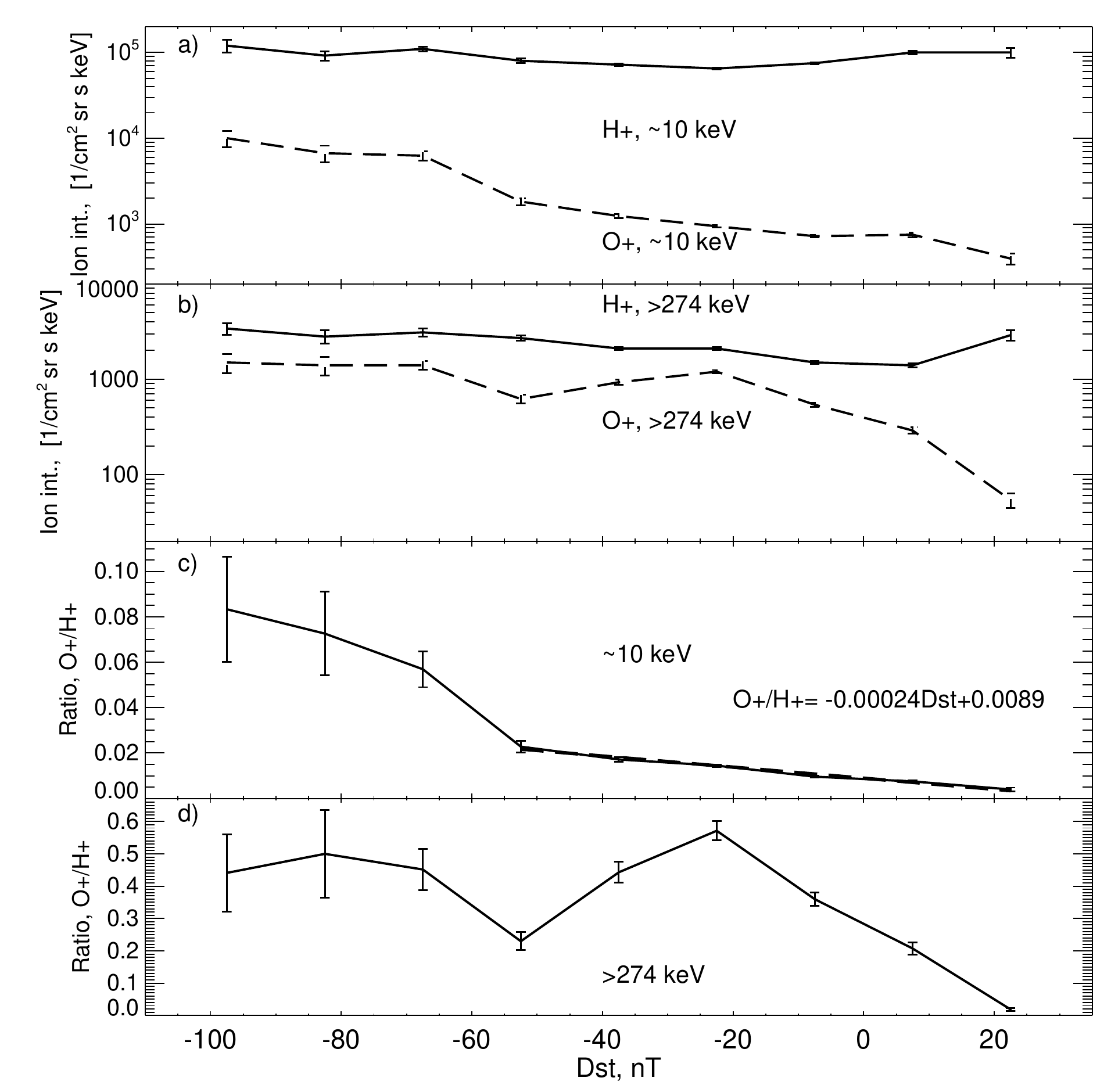}}
  \caption{a) Dependencies on Dst index of (a) proton and oxygen
intensities at $\sim$10 keV, (b) proton and oxygen intensities for
$>$274 keV, (c) ratio O$^+$/H$^+$ for intensities at $\sim$10 keV
and (d) ratio O$^+$/H$^+$ for intensities at $>$274 keV. The
dashed line in panel 3 indicates a linear fit to the ratio values. Error
bars indicate 99$\%$ confidence intervals. \label{fig:dst}}
\end{figure}

The response of the ion intensities to the changes in the Dst index is shown
in Figure \ref{fig:dst} and summarized in Table \ref{tab:dstdep}. The
intensity of protons at $\sim$10 keV and at $>$274 keV increases
approximately in the same manner with decrease of Dst index down to -100 nT
(the maximum intensity is $\sim$2 times higher than the minimum intensity).
Oxygen intensities at both energies show more dramatic changes with Dst
index. Intensities are up to 30 times higher between the minimum to the
maximum. However, during weak (Dst below -30 nT) and moderate (Dst below -50
nT) magnetic storms, the ratio O$^+$/H$^+$ is affected more strongly at $\sim$10
keV. Since the $\sim$10 keV O$^+$ intensity grows significantly (5 times) at
Dst from -30 to -100 nT, compared to no change of O$^+$/H$^+$ ratio at $>$274
keV. At the higher energies, the intensities of both species grow together
during weak and moderate storms, therefore, the O$^+$/H$^+$ ratio does not
clearly change. This agrees with \citet{Ono09} who found that during
dipolarization associated with substorms the energy density of energetic
oxygen does not always increase. However, we see a significant growth of the
$>$274 keV oxygen, and of O$^+$/H$^+$ ratio during the growth phase, from 15
to -30 nT. Therefore, a significant amount of $>$274 keV oxygen is
effectively accelerated during a storm growth phase.

\begin{table}
\caption{Response of ion intensity to changes in the Dst index.} \label{tab:dstdep}
\begin{tabular}{clllll}
\hline
Particles&Compression\textsuperscript{a}&Growth phase\textsuperscript{b}&Weak-moderate\textsuperscript{c}&Quiet-Storm time\textsuperscript{d}&Min-Max\textsuperscript{e} \\
&Dst$>$15 nT\textsuperscript{f}&-30$<$Dst$<$15 nT&-105$<$Dst$<$-30 nT&-105$<$Dst$<$0 nT & \\
\hline
H$^+$ $\sim$10 keV&no change&no change&+1.6x\textsuperscript{g}&+1.4x&+2x            \\
O$^+$ $\sim$10 keV&-2x\textsuperscript{g}&+1.5x&+8x&+14x&+13x                \\
O$^+$/H$^+$ $\sim$10 keV&-2x&+1.7x&+5x&+10x&+22x             \\
H$^+$ $>$274 keV&+2x&+1.5x&+1.6x&+2.3x&+2.4x                \\
O$^+$ $>$274 keV&-8x&+3x&+1.6x&+3.6x&+30x                   \\
O$^+$/H$^+$ $>$274 keV&-16x&+2x&no change&+1.6x&+30x    \\
\hline  \\
\end{tabular}
{\textsuperscript{a} Difference between ion intensities at $\sim$22.5 nT and $\sim$0 nT (here and further, the mean values of the corresponding bins), namely during magnetospheric compression.}\\
{\textsuperscript{b} Difference between ion intensities at $\sim$0 nT and $\sim$-22.5 nT, namely between beginning and end of the growth phase.}\\
{\textsuperscript{c} Difference between ion intensities at $\sim$-37.5 nT and $\sim$-97.5 nT, namely between weak and moderate storms.}\\
{\textsuperscript{d} Difference between ion intensities at $\sim$0 nT and $\sim$-97.5 nT, namely between quiet and storm time.}\\
{\textsuperscript{e} Difference between minimum and maximum ion intensities in Figure \ref{fig:dst}.}\\
{\textsuperscript{f} This value and corresponding values in the next columns are the thresholds of the ranges.}\\
{\textsuperscript{g} ``-*x'' is decrease in * times and ``+*x'' is increase in * times.}\\

\end{table}

At positive Dst$>$15 nT the intensity of $>$274 keV protons is clearly
higher compared to that at Dst$\simeq 0$. This is expected as positive Dst values are
associated with a compression of the magnetopause which can lead to a drift
shell displacement. An enhanced intensity of $\sim$10 keV H$^+$ is observed
for positive values of Dst. The effect is not observed in the O$^+$
intensity, though.

Based on the observations, we can define linear functional dependence between
the O$^+$/H$^+$ ratio and disturbance level (from -60 to 30 nT):
$$O^+/\mbox{H}^+=-9.7\cdot 10^{-4} \times \mbox{Dst[nT]}+3.5\cdot 10^{-2}.$$ The dependency
is constructed using a linear least squares fit.

We find a weak correlation between proton intensity (both at $\sim$10 keV and
$>$274 keV) and Dst. This is consistent with results reported by e.g.
\citet{nose2009a,Mouikis10}. Also consistent with our results,
\citet{Mouikis10} found significantly stronger correlation of oxygen
intensities and geomagnetic activity in the magnetotail region.

\subsubsection{Response to magnetotail processes}\label{sec:AE}

Processes in the magnetotail due to e.g. magnetospheric substorms and bursty
bulk flows causing enhanced field-aligned currents leading to auroral
electrojet activity reflected in the AE index.

\begin{figure}
 {\includegraphics[width=150mm]{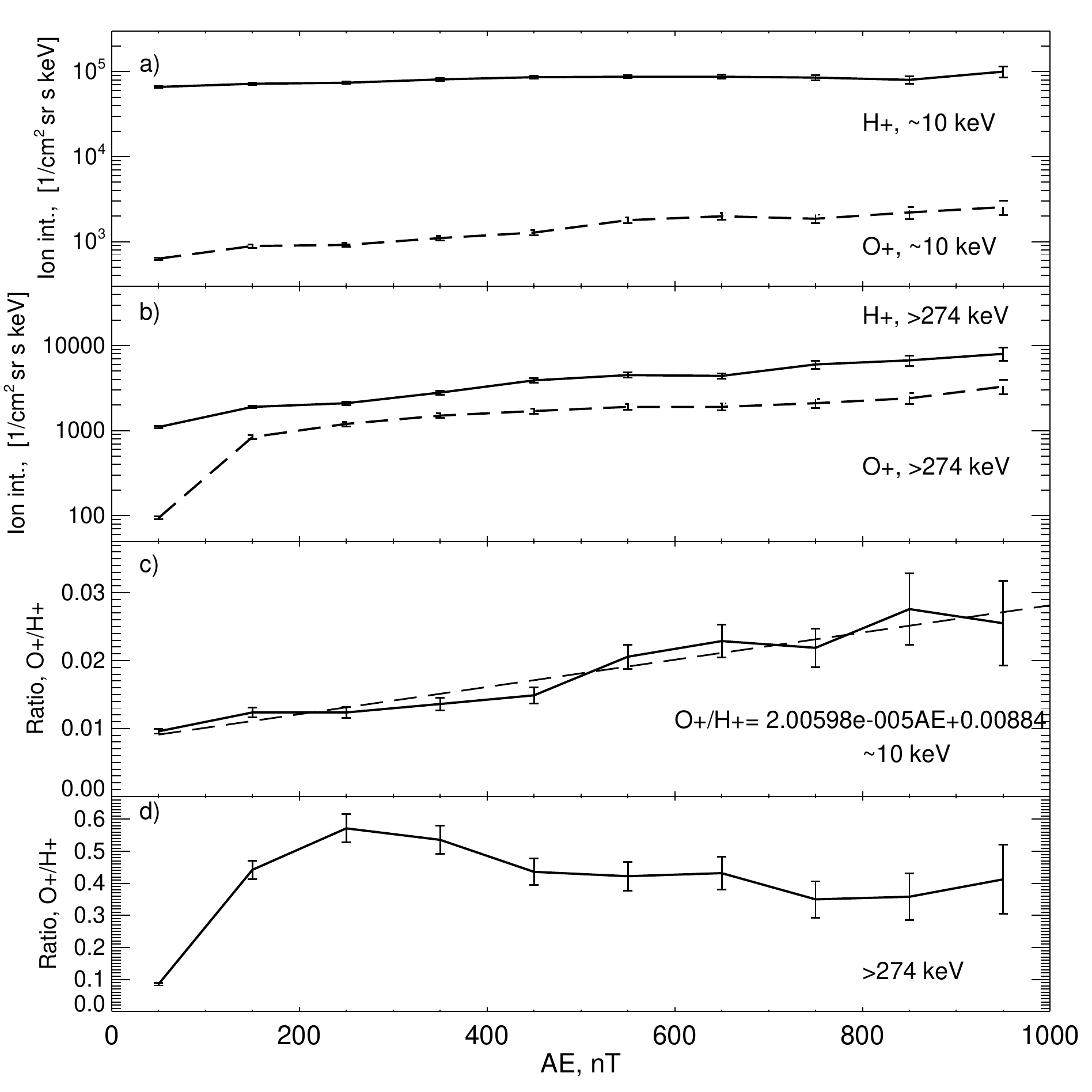}}
  \caption{Dependencies on AE index of (a) proton and oxygen
intensities at $\sim$10 keV, (b) proton and oxygen intensities at
$>$274 keV, (c) ratio O$^+$/H$^+$ for intensities at $\sim$10 keV
and (d) ratio O$^+$/H$^+$ for intensities at $>$274 keV. The
dashed line in panel 3 indicates a linear fit to the ratio values. Error
bars indicate 99$\%$ confidence intervals. \label{fig:ae}}
\end{figure}

Figure \ref{fig:ae} shows the response in H$^+$ and O$^+$ intensities to these processes.
The increase in the AE index is associated with an exponential growth by a
factor 4 (from the minimum to the maximum) of the $\sim$10 keV oxygen
intensities. The $\sim$10 keV protons only show a slight increase with AE index,
(maximum value $\sim$1.5 time higher than the lowest value). As above, we can now
establish an empirical functional relation between $\mbox{O}^+/\mbox{H}^+$ ratio and AE
index for $\sim$10 keV ions
$$\mbox{O}^+/\mbox{H}^+=8\cdot 10^{-5}\times \mbox{AE[nT]}+3.5\cdot 10^{-2}.$$
The increase of the $\mbox{O}^+/\mbox{H}^+$ ratio at $\sim$10 keV energies
means that the oxygen is accelerated up to these energies more effectively
than protons, although sources of their acceleration have to be different.
The study by \citet{Ono09} shows similar results: acceleration of O$^+$ at
energies of 9 to 36 keV/e during substorms (by electric fields induced by
magnetic field fluctuations whose frequencies are close to the
gyrofrequencies of ions) is more effective than of H$^+$. Event studies by
\citet{Zong08,Zong09} have shown that bursty bulk flows can be responsible
for supply of the ionospheric oxygen into the near-Earth magnetosphere.

Proton intensities at $>$274 keV show an exponential increase with AE.
Intensities are $\sim$8 times higher for high AE values. A similar dependence
between AE and $>$274 keV oxygen intensities are also observed, although the
slope has a break at 150-250 nT (strong growth before and moderate after).
The highest intensities of $>$274 keV oxygen, at high AE values are about 33
times higher than the lowest intensities.

The effect of substorms on the abundance of the O$^+$ ions relative to
protons is larger at $\sim$10 keV, since the O$^+$ intensity increases more
strongly than H$^+$. At higher energies, the intensities of the two species
seem to increase approximately in the same way starting from AE$\simeq$200
nT. This again agrees with results obtained by \citet{Ono09} who did not find
a clear correlation between substorm-associated dipolarizations and effective
O$^+$ acceleration for ions in the energy range 56-212 keV/e. However, our
results show a significant increase in intensities of $>$274 keV oxygen from
quiet (AE$\simeq$0-100 nT) to more disturbed times (AE$\simeq$150-250 nT).
Therefore, a significant amount of $>$274 keV oxygen is effectively
accelerated during weak/moderate substorms or during a substorm growth phase.

\subsection{Response to solar wind conditions}

\subsubsection{Response to IMF direction}\label{sec:CA}

\begin{figure}
 {\includegraphics[width=150mm]{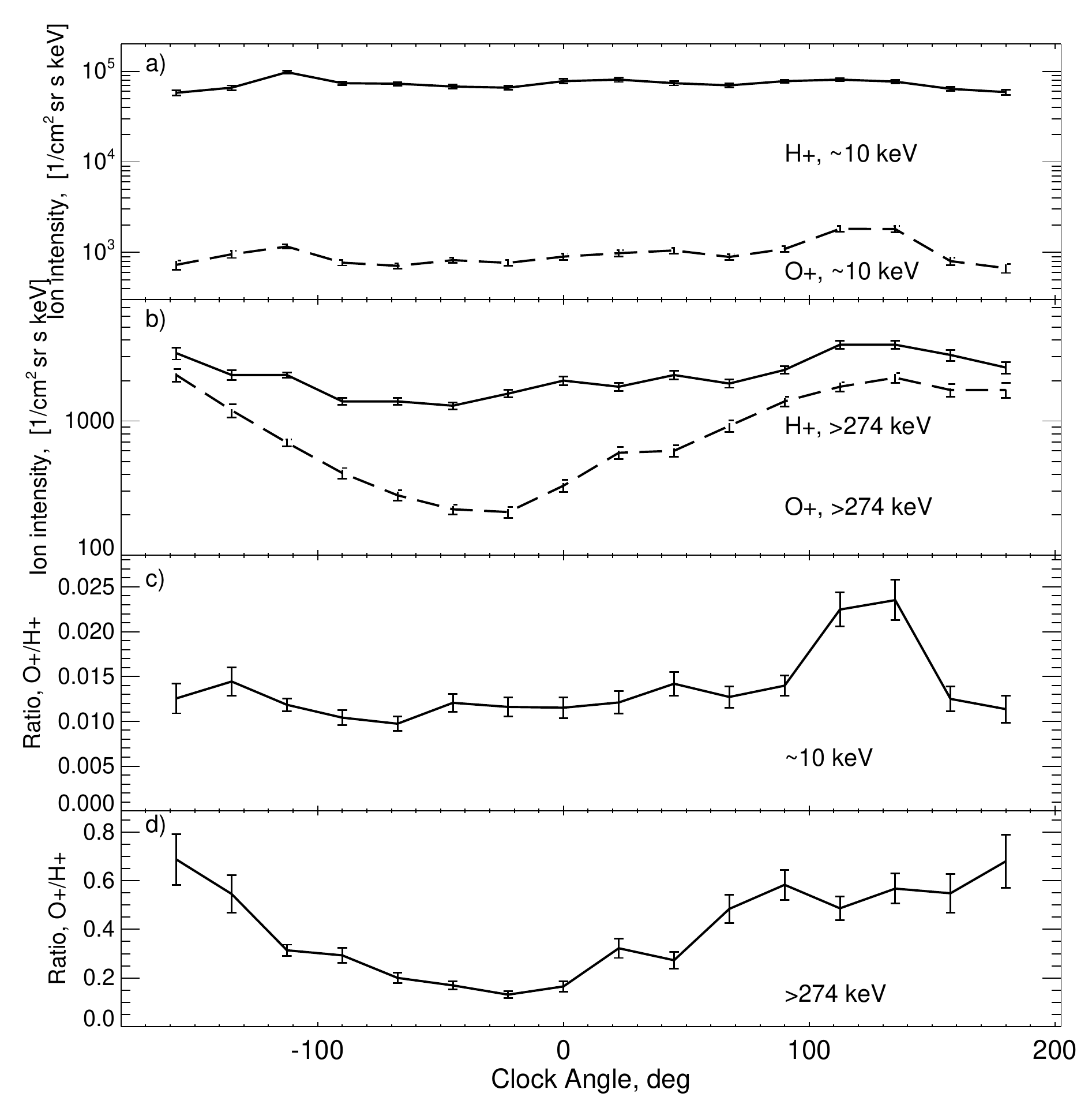}}
  \caption{
Dependencies on Clock Angle of
(a) proton and oxygen intensities at $\sim$10 keV,
(b) proton and oxygen intensities at $>$274 keV,
(c) ratio O$^+$/H$^+$ for intensities at $\sim$10 keV and
(d) ratio O$^+$/H$^+$ for intensities at $>$274 keV.
\label{fig:imf}}
\end{figure}

The response of the ion intensities to the changes in IMF direction is shown
in Figure \ref{fig:imf}. The IMF direction is represented by the clock angle
(CA) defined as the angle between $Z_{GSE}$ and the projection of IMF into
the $YZ_{GSE}$ plane.

When IMF Bz is negative (i.e. southward IMF), the coupling to the Earth's geomagnetic field is strongest.
Southward interplanetary magnetic conditions are typically associated with geomagnetic
storm activity.

In order to take into account that the changes in particle population do not immediately respond to
changes in IMF direction, we only use those measurements during periods where
IMF was reasonably stable for at least 30 minutes.

The intensity of $\sim$10 keV protons increases slightly at the IMF clock
angle, -115$^{\circ}$ in $\sim$1.5 times from the minimum to the maximum. For
the oxygen intensities the increases are observed at -115$^{\circ}$ and
125$^{\circ}$, in $\sim$2.5 times from the minimum to the maximum.

High energy particles (energy $>$274 keV) are much more sensitive to IMF
clock angle variations, especially oxygen ions. The oxygen and proton
intensities at $>$274 keV are a factor of 10 and 3 higher for southward IMF,
respectively. We find a strong positive correlation between the intensities
of $>$274 keV O$^+$, H$^+$, the O$^+$/H$^+$ ratio and the IMF direction. The energetic
ion intensities seem to be correlated with the``openness'' of the
magnetosphere or change of the clock angle towards the southward direction.
However, during the most southward IMF (and dawnward) the acceleration ceases
to be effective, the O$^+$/H$^+$ ratio does not show significant changes.

Since RAPID does not discriminate between charge states, oxygen of
solar wind origin (with higher charge states) might explain the strong
correlation with IMF clock angle. To check this, we compared our intensities
with corresponding intensities of high charge state oxygen, with clear solar
wind origin, measured by \citet{Kremser87} in the inner magnetosphere using
the AMPTE/CHEM data set. Their results indicate that intensities of high
charge state oxygen were about one order of magnitude lower than our RAPID
results. Hence, we think that our observations are primarily of ionospheric
origin. The effect of IMF orientation is significantly larger for the
 intensity ratio of the O$^+$ ions to protons at $>$274 keV than at $\sim$10
 keV. These all suggest a connection between the energy input to the
magnetosphere related to the change of clock angle and the energization of
O$^+$ to high energies ($>$274 keV).

\subsubsection{Solar wind density and pressure}\label{sec:SW}

\begin{figure}
{\includegraphics[width=150mm]{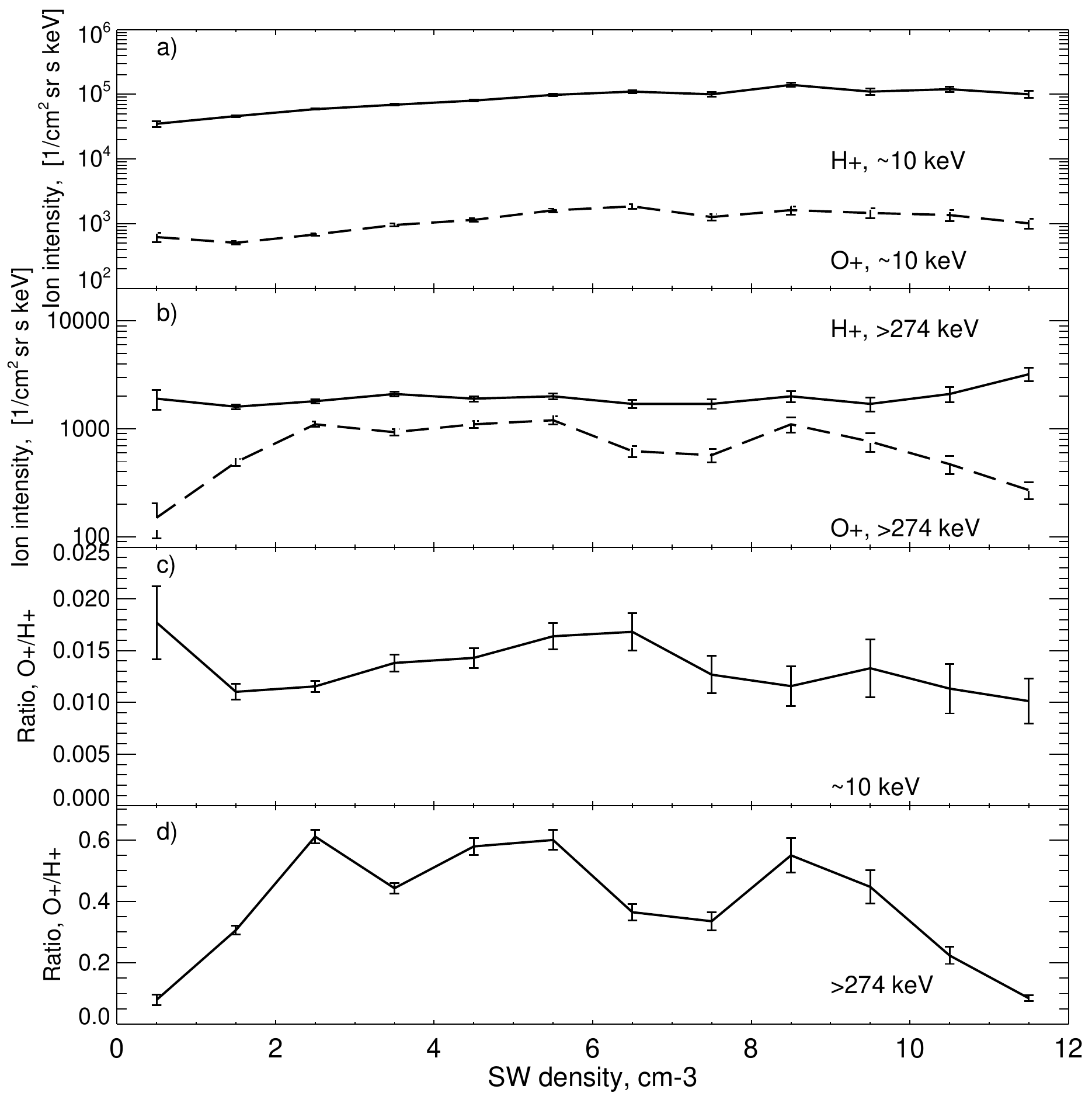}}
  \caption{
Dependencies on solar wind density of
(a) proton and oxygen intensities at $\sim$10 keV,
(b) proton and oxygen intensities at $>$274 keV,
(c) ratio O$^+$/H$^+$ for intensities at $\sim$10 keV and
(d) ratio O$^+$/H$^+$ for intensities at $>$274 keV.
\label{fig:swdensity}}
\end{figure}

\begin{figure}
  {\includegraphics[width=150mm]{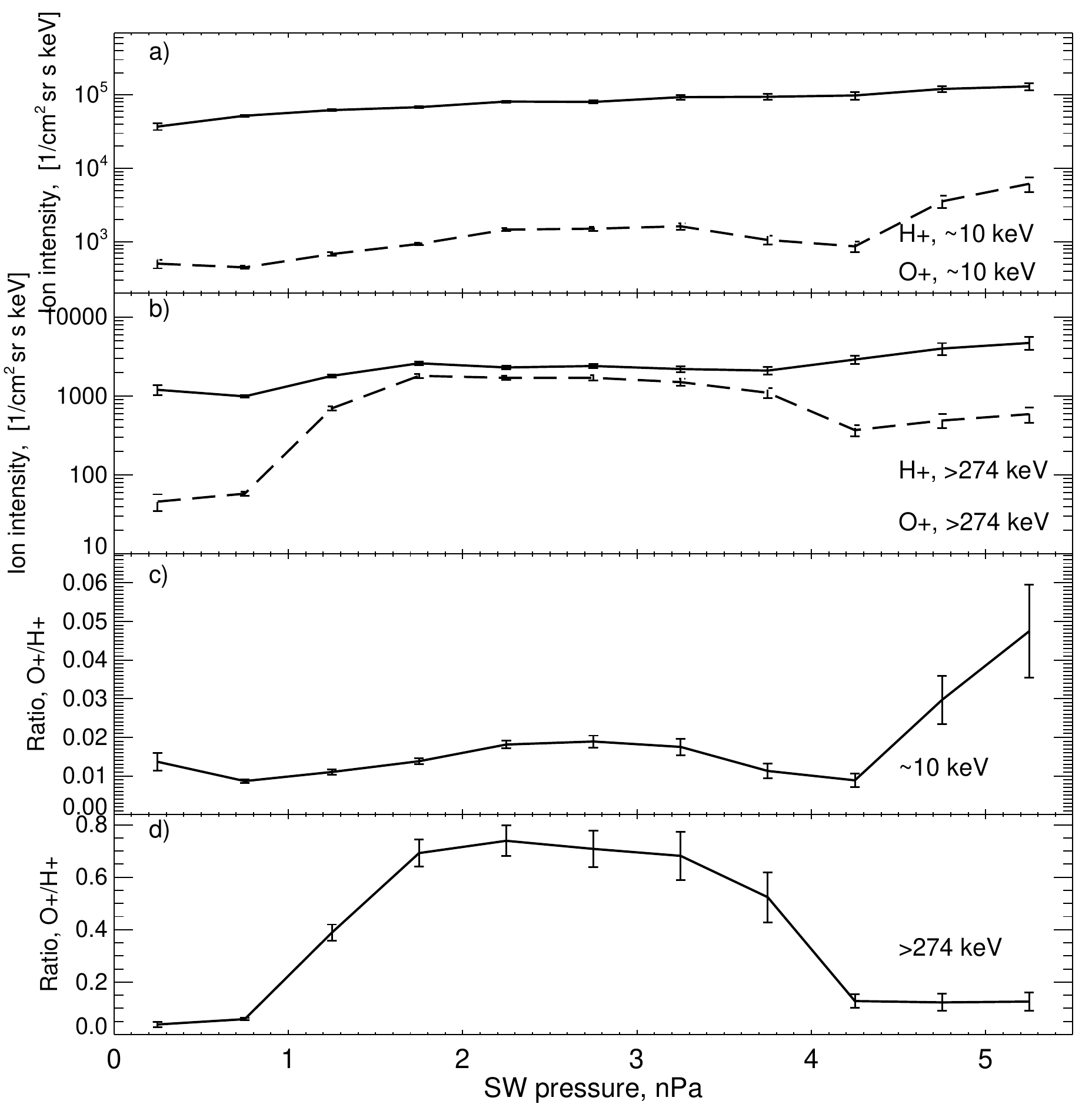}}
  \caption{
Dependencies on solar wind pressure of
(a) proton and oxygen intensities at $\sim$10 keV,
(b) proton and oxygen intensities at $>$274 keV,
(c) ratio O$^+$/H$^+$ for intensities at $\sim$10 keV and
(d) ratio O$^+$/H$^+$ for intensities at $>$274 keV.
\label{fig:swpressure}}
\end{figure}

The response of the ion intensities to changes in solar wind density and
pressure is shown in Figures \ref{fig:swdensity} and \ref{fig:swpressure},
respectively. Intensities of $\sim$10 keV oxygen and protons are
almost exponentially increasing with solar wind density, (by a factor of $\sim$4 for
H$^+$ and a factor of $\sim$3 for O$^+$. For energies $>$274 keV, we do
not see any clear dependence on the solar wind density.

The $\sim$10 keV protons and oxygen show almost exponential increase (by a
factor $\sim$ 4) with the solar wind dynamic pressure. The proton intensities
at $>$274 keV increase (by a factor of $\sim$ 5) with the solar wind pressure.
There is no clear general dependency of the $>$274 keV oxygen intensities on
the solar wind pressure. However, for $>$274 keV O$^+,$ we observe a $\sim$40
fold increase between $\sim$0.5 nPa to 1.75 nPa. At these solar wind pressure
values, the energetic oxygen is accelerated more effectively than energetic
protons. This is also supported by the fact that in $\sim$10 keV oxygen such
a tendency is not observed. From $\sim$1.75 nPa to 3nPa the intensity of the
$>$274 keV oxygen and protons becomes almost equal.

Statistically, the solar wind density strongly correlates with the plasma
sheet density \citep{Borov98c}. This is seen for the $\sim$10 keV protons.
For $\sim$10 keV protons we find a strong correlation between intensity and
the solar wind density and pressure.

The increase of $\sim$10 keV O$^+$ intensities is well correlated with the
increase of the solar wind dynamic pressure. This is in agreement with
earlier studies, e.g. by \citet{Cully03b,Nose03} who found that the
ionospheric outflow is strongly affected by the solar wind dynamic pressure.
Also in a study by \citet{Echer08}, it was shown that a shock compression can
lead to enhanced inflow of the oxygen ions into the near-Earth magnetosphere.
However, this can be also compression effects, see Section \ref{sec:swpdisc}.

Intensities of $\sim$10 keV oxygen ions show clear dependencies on solar wind
dynamic pressure, as this energy is closer to source energies. This is not
the same for energetic ions which are not closely related with outflow
energies and require stronger acceleration. The observed prominent increase
in the oxygen intensities for solar wind dynamic pressure up to 1.5 nPa is
more likely related to acceleration mechanisms due to the thinning of the
plasma sheet, see Section \ref{sec:swpdiscgr}. The largest increase in the
energetic particle intensities is seen during the transition period between
quiet and disturbed times.

\section{Discussion}
\label{sec:discussion}

\subsection{Comparison to other observations}\label{compar}

Our energy density results listed in Table \ref{tab:ratios} (see calculations
in Appendix, Equation \label{eq:ed}) are comparable to the observations in the near-Earth plasma
sheet made by e.g. \citet{gloeck85a}, \citet{hami88a} and \citet{nose2001a}.
We do not observe the dramatic differences between quiet and disturbed times
reported in these earlier studies, though. A possible reason is that we have
used median values rather than mean values, and thus avoid effects of
outliers and strongly scattered distributions. Also, the definitions of quiet
and disturbed times, are based on AE index and Dst indices and differs from
earlier studies. One could notice that the difference between quiet and
disturbed times for the $\sim$10 keV ions is stronger when using the Dst
index as definition. Taking even more negative values of Dst will lead to the
higher O$^+$/H$^+$ ratios. Therefore, the dynamics of the $\sim$10 keV
particles in the near-Earth magnetosphere is more strongly correlated with the Dst
index.

\begin{table}
\caption{$\mbox{O}^+/\mbox{H}^+$ ratios of energy density depending on the disturbance
level and the location.} \label{tab:ratios}
\begin{tabular}{cllll}
\hline
Satellite,&Energy range,& $\mbox{O}^+/\mbox{H}^+,$ Energy density& \\
instrument &keV &    Quiet time & Disturbed time \\
\hline
Cluster/RAPID   &274 keV - $\sim$955 keV &0.22$\pm$0.12\textsuperscript{a}, 0.72$\pm$0.44\textsuperscript{b}& 1.04$\pm$1.33\textsuperscript{a}, 1.11$\pm$1.35\textsuperscript{b}\\
Cluster/CIS    &$\sim$10 keV&0.038$\pm$0.0081\textsuperscript{a}, 0.034$\pm$0.0031\textsuperscript{b}& 0.083$\pm$0.038\textsuperscript{a}, 0.2$\pm$0.12\textsuperscript{b}\\
AMPTE/CCE\textsuperscript{c}& 1-310 keV&0.03 &0.34\\
AMPTE/CCE\textsuperscript{d}& 1-310 keV&0.01 &0.61\\
Geotail/EPIC\textsuperscript{e}& 9-210 keV&0.05-0.1&0.2-0.6\\
\hline  \\
\end{tabular}
{\textsuperscript{a} Quiet (AE$<$100 nT) and disturbed conditions (AE$>$300 nT) are based on the AE index.}\\
{\textsuperscript{b} Quiet (Dst$\sim$ 0 nT) and disturbed (Dst is between -100 and -30 nT) are determined based on the Dst index.}\\
{\textsuperscript{c} Measurements taken from \citet{gloeck85a}.}\\
{\textsuperscript{d} Measurements taken from \citet{hami88a}.}\\
{\textsuperscript{e} Measurements taken from \citet{nose2001a}.}\\
\end{table}

\subsection{Comparison to ionospheric outflow observations}

Ion outflow rates (at 0.01-17 keV) can vary by a factor of 30 between quiet
and very disturbed periods \citet{yau85}. Our result shows that the $>$274
keV oxygen intensity is about 33 times higher during very disturbed periods
than during quiet time (here AE index is used). For comparison, the $\sim$10
keV O$^+$ intensity increases a factor 4 between quiet and very disturbed
times. Acceleration of $\sim$10 keV oxygen seems to be more effective during
disturbed times, while for $>$274 keV oxygen acceleration primary occurs
during transitions from quiet to disturbed periods. Such slope
break between quiet and disturbed states is not observed in the ionospheric
O$^+$ outflow \citep{yau85,Cully03b}.

For protons, the intensities of $\sim$10 keV ions are only slightly different
during disturbed periods than during quiet times. The intensity of protons
with energies $>$274 keV increases by a factor of about 8 between quiet and
disturbed times. For comparison, for auroral and polar protons outflow an
increase of the order of 5 was reported by \citet{yau85}.

The ratio between quiet and disturbed times for the ion ionospheric outflow
is higher than those observed in the near-Earth magnetosphere at $\sim$10 keV
but similar to those observed at $>$274 keV. These appears to be a connection
between the energetic ion intensities and ion outflow that changes with
geomagnetic activity. The fact that the ratios between quiet and disturbed
times are higher for the more energetic and heavier particles means that the
ions are more effectively accelerated up to these energies. This is
consistent with explanation by \citet{moeb87} and many other studies thereafter,
that the acceleration process in the near-Earth magnetosphere is mass
dependent, as it is more effective for the heavier particles.

\subsection{Response of protons to geomagnetic activity and solar wind conditions}

As shown in Section \ref{sec:CA}, the intensity of $\sim$10 keV protons does
not seem to be directly affected by dayside reconnection. Response to IMF
direction would be expected, though, as it is likely to be associated with
dayside reconnection and then consequent transfer of the solar wind ions into
the magnetosphere and subsequent acceleration. By the same token, magnetic
storms and substorm activity are typically also related to the dayside
reconnection. In addition storms and substorms are typically associated with
enhanced outflow. As a consequence, more ions would be energized in the
magnetosphere. Still, the correlation between the $\sim$10 keV proton
intensities on one hand, and geomagnetic activity, solar wind dynamic
pressure and clock angle on the other hand is less pronounced than for the
$\sim$10 keV oxygen, see Sections \ref{sec:Dst}, \ref{sec:AE} and
\ref{sec:CA}.

These observations can be explained as follows. Heavier ions (O$^+$) require
additional forces such as electric fields, wave activity or stronger
gradients in pressure and temperature to escape from the ionosphere. These
processes are primarily driven by enhanced dayside reconnection and the
subsequent processes reflected by increased disturbance levels. Protons, on
the other hand, with their much lower escape energies, are not so dependent
on these external forces. Solar illumination alone suffices to extract a
significant number of light ions from the ionosphere. Therefore, the strong
dependence on the disturbance parameters of the $\sim$10 keV protons is not
observed.

\subsection{Changes during growth phase}\label{sec:swpdiscgr}
Why would we expect an increase in energetic oxygen intensities during the
transition period between quiet and disturbed times (substorm growth phase)?
The increased solar wind pressure (or loading of the energy during Dungey
cycle) can lead to the plasma sheet thinning \citep[e.g.,][]{Sauvaud96b} and
therefore to a more effective acceleration of the heavy ions
\citep[e.g.,][]{Ganguli95}. There are different regimes of particle motion in
the plasma sheet derived by \citet{Buechner86}. According to their work the
particles are accelerated most effectively when parameter $\kappa
=\sqrt\frac{r_{curv}}{r_g}\sim 1,$ where $r_{curv}$ is the radius of magnetic
field curvature and $r_g$ is ion gyroradius. Under these conditions the
particle trajectories become stochastic and particles are effectively
accelerated by quasi-stationary dawn-dusk electric fields. A decrease in
plasma sheet thickness, e.g. in response to an increase of the lobe magnetic
field, will lead to larger region where stochastic acceleration of energetic
ions can take place, as it will decrease the parameter $\kappa$ which is
quite large in the plasma sheet for solar wind protons \citep{Ashour09}.

\subsection{Changes with solar wind pressure}\label{sec:swpdisc}

Increased solar wind dynamic pressure leads to a compression of the
magnetosphere and the same total quantity of particles have to be distributed
over a smaller volume. Consequently, the particle density/intensity will be
higher and this will lead to the observed correlation. The
observed differences in behavior between $\sim$10 keV (where a correlation is
evident) and $>$ 274 keV ions (no clear correlation with dynamic solar wind
pressure) can be explained as follows: the plasma sheet may become thinner as result of
very strong solar wind pressure, and therefore, the gyroradius versus curvature
radius effect discussed above may be relevant also here.

On other hand, the correlation with the solar wind dynamic pressure can be
related to the Earth's ionosphere which may provide heavier ions to the magnetotail
during enhanced solar wind dynamic pressure, see
\citet{Cully03b,Nose03,Echer08}.

\section{Summary}
\label{sec:summary}

 Based on seven years of ion composition data from Cluster observations at
 radial distances -10 \,R$_{\rm E}$\ $<X_{GSE}<$ 10 \,R$_{\rm E}$, we find the following:

 (1) H$^+$ intensities at $\sim$10 keV show only a slight correlation with
geomagnetic conditions (Dst and AE indices) and interplanetary magnetic field
orientation. Solar wind dynamic pressure and density (i.e. effects of
magnetospheric compression) seem to have a larger effect on the $\sim$10 keV H$^+$ intensities.

(2) O$^+$ ion intensities at $\sim$10 keV are more affected by geomagnetic
storms and substorms than $>$274 keV O$^+$ ion intensities, than their
corresponding hydrogen counterparts. The $>$274 keV O$^+$
energization/acceleration seem to be strongest during the transition period
from quiet to disturbed times, i.e., during growth phases rather than during
the disturbed phases itself. The $\sim$10 keV ions do not reveal such a
dependence.

 (3) We find a strong positive correlation between the flux of $>$274 keV
O$^+$, H$^+$, the O$^+$/H$^+$ ratio and the IMF direction. This demonstrates a connection between energy input to
the magnetosphere and effective energization of energetic ions.

(4) The intensity ratio between quiet and disturbed times for the ionospheric ion outflow
is similar to those observed in the near-Earth magnetosphere at $>$274 keV.
Therefore, the observed increase of the energetic ion intensity during disturbed time is
not only due to a more effective acceleration but also due to enhanced ion outflow.

(5) Our results seem to confirm the conclusion of \citet{moeb87} and many other
studies afterwards, in that acceleration processes in the near-Earth
magnetosphere are mass dependent, because it is more effective for the heavier
ions (O$^+$/H$^+$ ratio higher).

\appendix
\section{Construction of
RAPID O$^+$/H$^+$ ratio}\label{app}
We shall refer to the corresponding differential particle
fluxes, $j_i,$ from the eight energy channels, ($E_n^{H^+},$ $E_n^{O^+}$), of the RAPID
instrument as $j_1^{H^+}$ to $j_8^{H^+}$ and $j_1^{O^+}$ to $j_8^{O^+}$,
respectively (see Table \ref{tab:energies}).

The lowest energy channel for oxygen, $E_1^{O^+}$ (with energy range 84--274
keV), is contaminated by protons, and cannot be used \citep{Daly10}. The
count rates in energy channels above 1 MeV (i.e., $E_6,$ $E_7$ and $E_8$) are
usually extremely low for both species. Useful O$^+$ channels are therefore
energy channels $E_2^{O^+}-E_5^{O^+}$, covering energies from 274 to
$\sim$955 keV. For protons, corresponding to this energy range, parts of
$E_4^{H^+}$ and $E_5^{H^+}$ have to be used. We have decided to use both these
H$^+$ channels rather than only one $E_5^{H^+}$ and the part of the
$E_2^{O^+},$ respectively, for better statistics. To construct O$^+$ and
H$^+$ energy bins with matching energies, we proceed as follows:

\begin{figure}
{\includegraphics[width=150mm]{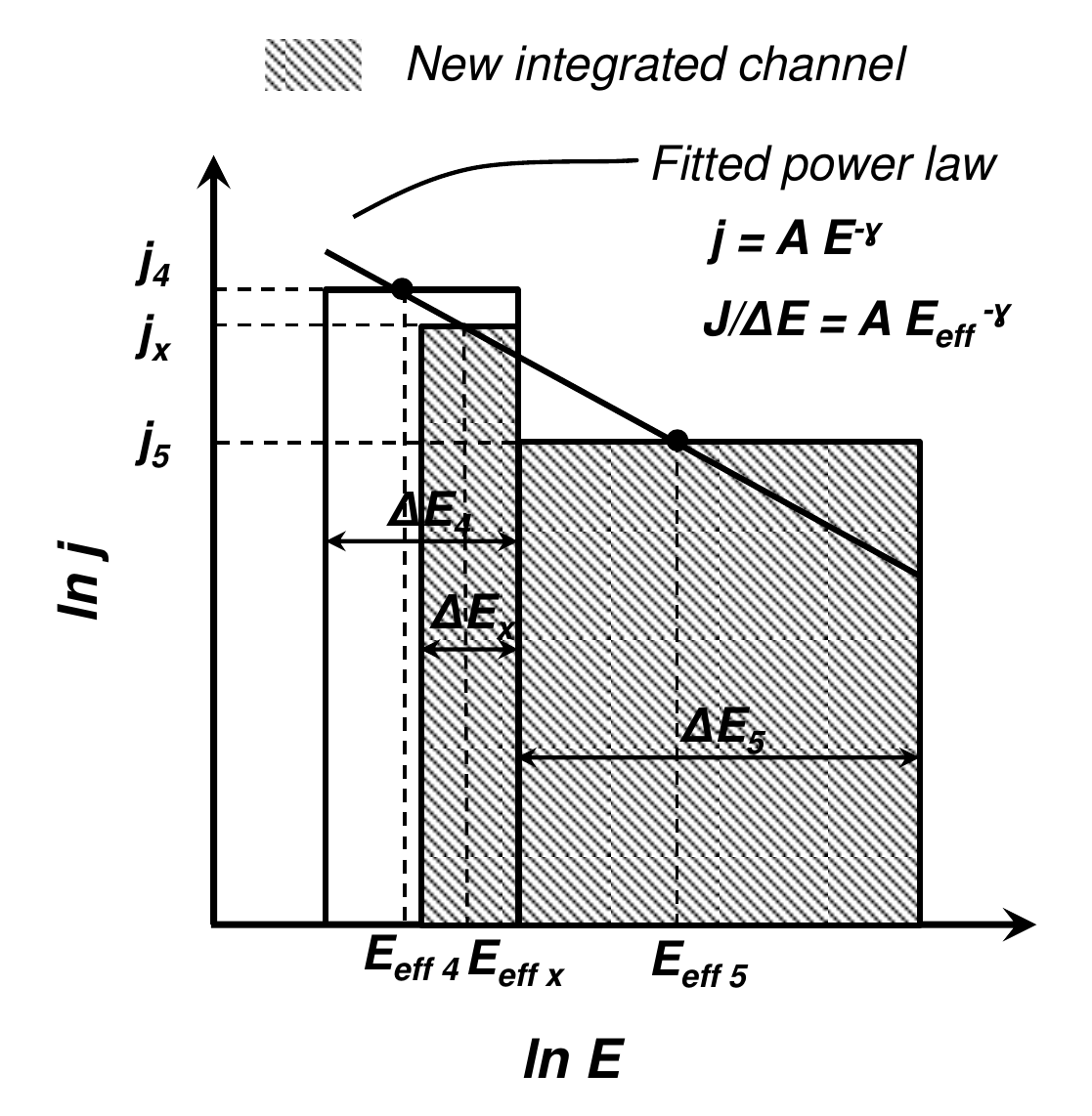}}
  \caption{ Bringing RAPID H$^+$ channels to the same energy range as O$^+$
channels: the differential flux in channels $E_4^{H^+}$ and $E_5^{H^+}$ is
fitted to a power law, that fit is used to determine the differential flux,
$j_x,$ in the virtual channel $E_x$. $E_{eff}$ is the energy at which the
spectrum with given $A$ and $\gamma$ has the differential flux equal
$J/\Delta E,$ where $J$ is the integrated intensity. \label{fig:sketch}}
\end{figure}

\begin{enumerate}
\item
Determine the integrated oxygen intensity, $J^{O^+},$ at energies from
274 up to $\sim$955 keV. The integrated intensity is determined as the
differential fluxes, $j_i^{O^+}$ multiplied by respective
$\Delta E_i^{O^+}$: $$J^{O^+} = (j_2^{O^+}\cdot \Delta E_2^{O^+} + j_3^{O^+}\cdot
\Delta E_3^{O^+} + j_4^{O^+}\cdot \Delta E_4^{O^+} + j_5^{O^+}\cdot
\Delta E_5^{O^+}) [1/(cm^3\cdot sr\cdot s)];$$
\item
Obtain spectral index $\gamma,$ in order to cut the part of the $E_4^{H^+},$
see Figure \ref{fig:sketch} for illustration. For this we use fluxes
from two adjacent energy channels ($E_4^{H^+}$ and $E_5^{H^+}$) and their
effective energies, $E_{eff}$, the energy at which the spectrum, $A\cdot
E_{eff}^{-\gamma},$ with given $A$ and $\gamma$ has the differential flux
equal $J/\Delta E$: $$ \gamma = \ln(j_5^{H^+}/j_4^{H^+}) /
\ln(E_{eff 4}^{H^+}/E_{eff 5}^{H^+}), $$ where $E_{eff}$ in this case is
calculated as the geometric mean energy of the respective energy
channels. See more details on these calculations in the RAPID Calibration
Report \citep{rap:calrep}, chapter B.1;
\item
Determine $j_x^{H^+}$ --- i.e., the H$^+$ differential flux at the
virtual energy $E_x^{H^+}$, 274-374 keV (which matches the O$^+$ energy
channel), with effective energy $E_{eff x}^{H^+} = \sqrt{274\cdot374} $
keV and determined $\gamma$, see Figure \ref{fig:sketch}:
$$   j_x^{H^+} = j_4^{H^+} \cdot e^{-\gamma \cdot
ln(E_{eff x}^{H^+}/E_{eff 4}^{H^+})};$$
\item
Determine integrated proton intensity, $J^{H^+},$ at energies from 274 up to 962
keV:
$J^{H^+} = j_x^{H^+}\cdot \Delta E_x^{H^+} + j_5^{H^+}\cdot \Delta
E_5^{H^+}.$ With this procedure, we establish a virtual proton energy channel
with the same lower energy threshold as for the oxygen. The slightly different
upper energy level (962 and 948 keV, respectively) does not play any role
here;
\item Define the intensity ratio:
$\frac{\mbox{O}^+}{\mbox{H}^+} = \frac{J^{O^+}}{J^{H^+}}$
\end{enumerate}

In order to compare our results with other observations we need to use the
energy density ratio. To establish the energy density ratio, we assume that the relative geometric factors are correct for
both species:
\begin{equation}
\frac{\mbox{O}^+}{\mbox{H}^+} = \frac{\pi\sqrt{2m_O}\cdot\sqrt{E_{eff}}\cdot J^{O^+}\Delta E} {\pi\sqrt{2m_H}\cdot\sqrt{E_{eff}}\cdot J^{H^+}\Delta E}=\frac{4\cdot
J^{O^+}}{J^{H^+}}, \label{eq:ed}
\end{equation}
where O$^+=\sqrt{m_O}\cdot\sqrt{E_{eff}}\cdot J^{O^+}$ and $H^+=\sqrt{m_H}\cdot\sqrt{E_{eff}}\cdot J^{H^+}.$

This calculation is based on the assumption that the effective energy is equal
to the geometric mean of the corresponding energy thresholds. However, in our case the width of the energetic
channel is quite large and this will lead to the deviation of the energy
density from the value calculated using the effective energy as the geometric
mean. The way to calculate this deviation one can find in the the RAPID
Calibration Report \citep{rap:calrep}, chapter B.3. The deviation is
estimated to be $\sim$65$\%$ and error bar $\pm$30$\%$ from the value of the
energy density calculated in Equation \ref{eq:ed}. For these calculations the
typical range of $\gamma$ values derived from our data base were taken: for
O$^+$ $\gamma=$2--3.5 and for H$^+$ $\gamma=$3.5--6.5. The statistical errors of the energy density
and the error due to the large width of the energy channels are added in this
case.

\begin{acknowledgments}
We thank the Deutsches Zentrum f{\"u}r Luft und Raumfahrt (DLR) for
supporting the RAPID instrument at MPS under grant 50 OC 1101. The
authors are grateful for the use of the OMNI data base providing AE,
Dst and solar wind parameters. E. Grigorenko thanks grants RFBR
Nr. 10-02-00135; 10-02-93114; 12-02-92614 and grant of Leading Scientific
Schools HIII-623.2012.2.
\end{acknowledgments}

\bibliographystyle{agu04}

\end{article}

\end{document}